\documentclass[prd,
preprint,nofootinbib%
 ,secnumarabic%
,amssymb, amsmath,mathrsfs,nobibnotes,aps,11pt]{revtex4}
\usepackage{epsfig}
\usepackage{color}
\usepackage{ulem}
\usepackage{graphicx}
\usepackage{float}
\usepackage[caption=false]{subfig}
\usepackage{braket}
\usepackage{slashed}
\newcommand{\M}{{\cal M}}
\usepackage{tabularx}
\usepackage[colorlinks=true,citecolor=blue,linkcolor=blue]{hyperref}%
\begin{document}
% \title{VDM photon energy distribution}
% \maketitle
%%%%%%%%%%%%%%%%%%%%%%%%%%%%%%%%%%%%%%%%%%%%%%%%%%%%%%%%%%%%%%%%%%%%%%%%%%%%%%%%%%%%%%%%%%%%%%%%%%%%%%%

\title{Vector dark matter annihilation with internal bremsstrahlung}

\author{Gulab Bambhaniya$^{1,}$\footnote{ gulab@prl.res.in}}
\author{Jason Kumar$^{2,}$\footnote{jkumar@hawaii.edu}}
\author{\mbox{Danny Marfatia}$^{2,}$\footnote{ dmarf8@hawaii.edu }}
\author{Alekha C. Nayak$^{3,}$\footnote{ acnayak@iitk.ac.in}}
\author{Gaurav Tomar$^{1,}$\footnote{ tomar@prl.res.in}}
\affiliation{$^1$Physical Research Laboratory, Ahmedabad 380009, India }
\affiliation{$^2$Department of Physics \& Astronomy, University of Hawaii, Honolulu, HI 96822, USA}
\affiliation{$^3$Department of Physics, Indian Institute of Technology, Kanpur 208016, India}
%\author{Gulab Bambhaniya \footnote{Email: gulab@prl.res.in}}
%\affiliation{Physical Research Laboratory, Ahmedabad 380009, India }
%\author{Jason Kumar \footnote{Email: jkumar@hawaii.edu}}
%\affiliation{Department of Physics \& Astronomy, University of Hawaii, Honolulu, HI 96822, USA.}
%\author{Danny Marfatia \footnote{Email: dmarf8@hawaii.edu }}
%\affiliation{Department of Physics \& Astronomy, University of Hawaii, Honolulu, HI 96822, USA.}
%\author{Gaurav Tomar \footnote{Email: tomar@prl.res.in}}
%\affiliation{Physical Research Laboratory, Ahmedabad 380009, India }
%\author{Alekha C. Nayak \footnote{Email: acnayak@iitk.ac.in}}
%\affiliation{Department of Physics, Indian institute of Technology, Kanpur 208016, India}
\def\be{\begin{equation}}
\def\ee{\end{equation}}
\def\al{\alpha}
\def\bea{\begin{eqnarray}}
\def\eea{\end{eqnarray}}
% \author{Tanushree Basak$^{1,}$\footnote{Email: tanu@prl.res.in}}
% \author{Subhendra Mohanty$^{1,}$\footnote{Email: mohanty@prl.res.in}}
% \author{Gaurav Tomar$^{1,2,}$\footnote{Email: tomar@prl.res.in}}
% \affiliation{$^1$Physical Research Laboratory, Ahmedabad 380009, India }
% \affiliation{$^2$Indian Institute of Technology, Gandhinagar 382424, India }
% \def\be{\begin{equation}}
% \def\ee{\end{equation}}
% \def\al{\alpha}
% \def\bea{\begin{eqnarray}}
% \def\eea{\end{eqnarray}}
%%%%%%%%%%%%%%%%%%%%%%%%%%%%%%%%%%%%%%%%%%%%%%%%%%%%%%%%%%%%%%%%%%%%%%%%%%%%%%%%%%%%%%%%%%%%%%%%%%%%%%%%%%%%%%%

\begin{abstract}
We consider scenarios in which the annihilation of self-conjugate spin-1 dark matter to a Standard Model fermion-antifermion 
final state is chirality suppressed, but where this suppression can be lifted by the emission of an 
additional photon via internal bremsstrahlung.  We find that this scenario can only arise if the 
initial dark matter state is polarized, which can occur in the context of self-interacting dark matter.  In particular, 
this is possible if the dark matter pair forms a bound state that decays to its ground state 
before the constituents annihilate.  We show that the shape of the resulting photon spectrum is the same as for
self-conjugate spin-0 and spin-1/2 dark matter, but the normalization is less heavily suppressed in the limit of heavy mediators.  
%This opens up new windows in parameter space wherein internal bremsstrahlung can provide an observable indirect detection signature for dark matter annihilation, while remaining consistent with LHC constraints on the mediator mass.   
\end{abstract}
\maketitle

%%%%%%%%%%%%%%%%%%%%%%%%%%%%%%%%%%%%%%%%%%%%%%%%%%%%%%%%%%%%%%%%%%%%%%%%%%%%%%%%%%%%%%%%%%%%%%%%%%%%%%%%%%%%%%%%%
\section{Introduction}

It is well understood that dark matter (DM) annihilation or
decay to a Standard Model (SM) fermion-antifermion pair $\bar f f$ can be chirality suppressed.
Then, the dominant indirect detection process in the current epoch may instead involve the internal bremsstrahlung (IB)
of an additional gauge boson (i.e., $\bar f f \gamma, \bar f f Z, \bar f f g, \bar f f' W^{\pm}$ final states)~\cite{Bergstrom:1989jr,Flores:1989ru,Bringmann:2007nk,Barger:2009xe,Kachelriess:2009zy, Ciafaloni2011,Bell:2011if,Barger2012,Garny2012,DeSimone:2013gj,Giacchino:2014moa,Bringmann:2015cpa}.  These processes
may not only dominate the annihilation/decay rate, but may also yield a hard boson spectrum which can be more
easily distinguished from background.  As a result, these internal bremsstrahlung processes have been well studied
for the case of spin-0 or spin-1/2 dark matter.  In this Letter, we discuss a case that has not been considered so far:
chirality suppression of spin-1 dark matter annihilation lifted by
internal bremsstrahlung.

The annihilation of spin-1 dark matter particles $B$ to a fermion-antifermion pair has been studied in several specific models \cite{Cheng2002,Cheng:2003ju, Farzan2012, Yu2014}, and in these cases it is found
that there is no chirality suppression, implying that the two-body final state is dominant.  We point
out that this unsuppressed contribution arises from the $J=2$ $s$-wave initial state.  If the DM initial state
is unpolarized, then there is indeed no way to avoid this unsuppressed contribution to the $2 \rightarrow 2$
annihilation cross section.  But if the DM initial state is polarized, and the $J=2$ initial state is
projected out, then the $s$-wave $BB \rightarrow \bar f f$ matrix element will be chirality suppressed, and the internal
bremsstrahlung process will dominate.  This scenario can be realized in a simple model in which the annihilation occurs
through the formation of a $BB$ bound state, which decays to its ground state before the two constituents
annihilate.
If the ground state is not $J=2$, then the branching fraction for decay to the
$\bar f f$ final state will be chirality suppressed, and the primary bound state decay channel will be to a three-body
final state.

We focus on the case in which internal bremsstrahlung involves the emission of a photon ($\bar f f \gamma$).
There are general arguments which show that for self-conjugate spin-0 or spin-1/2 dark matter,
the photon spectrum adopts a common universal form which depends only on $r$, the ratio of the
mass of the mediating particle ($m_\Psi$) to the mass of the dark matter.  We will show that this argument
generalizes to the case of spin-1 dark matter with one key difference:
for spin-0 or spin-1/2 dark matter, the annihilation matrix element necessarily scales as $m_\Psi^{-4}$ in the $r \gg 1$ limit,
while for spin-1 dark matter, the matrix element only scales as $m_\Psi^{-2}$.

The structure of this paper is as follows.  In Section 2 we review the general arguments that underly the chirality suppression of dark matter annihilation 
to the $\bar f f$ final state and apply these arguments to the case of 
spin-1 dark matter, inferring that IB is only relevant for a polarized initial state.  In section 3, we 
present the IB photon spectrum for the case of spin-1 dark matter, and demonstrate that its shape 
is necessarily the same as in the spin-0 and spin-1/2 cases. We also provide a physical realization of internal bremsstrahlung as the
dominant annihilation channel in terms of the decay of a dark matter bound state.  In section 4, 
we conclude with a discussion of our results.

\section{Chirality suppression in vector dark matter annihilation}

Chirality suppression of the cross section for $s$-wave dark matter annihilation to $\bar f f$ arises
for spin-0 or spin-1/2 dark matter if the dark matter particle is self-conjugate (i.e., the particle is its own antiparticle) and if minimal flavor violation applies.  One can understand  this result from general
principles; see, for example, Ref.~\cite{Kumar2013}.  If the dark matter particle is self-conjugate, then the initial state consists of two identical particles,
and must be even under charge conjugation.  For an $s$-wave ($L=0$) initial state, this implies that
$S$ is even, which in turn implies $J=0$.  The final state $\bar f f$ pair must then have $J_z=0$, where
the $z$-axis is taken to be the direction of motion of the outgoing particles.  Since $L_z$ vanishes along
the direction of motion, the final state particle and antiparticle must have the same helicity, and thus arise from
different Weyl spinors.  The matrix element thus violates SM flavor symmetries, and is necessarily
chirality suppressed by a factor $(m_f / m_X)^2$ under the assumption of minimal flavor violation, where $m_X$ is the mass of the 
dark matter.

One can also see why the annihilation of unpolarized spin-1 dark matter does not exhibit chirality suppression.
If the initial state consists of two identical real spin-1 particles, then it can also be in an $L=0$, $S=2$, $J=2$ state.
In this case, the final state need not mix different Weyl spinors, and thus the matrix element need not be chirality suppressed.

These results also follow from an analysis of the 4-point effective contact operators that can mediate dark
matter annihilation to a fermion-antifermion pair~\cite{Kumar2013,Kumar:2015wya}.  In particular, for spin-0 or spin-1/2
dark matter, one finds that there exists no operator of dimension $\leq 6$ which has a nontrivial matrix element with
an $s$-wave initial state of identical dark matter particles, and which also does not mix SM 
Weyl spinors.  But for spin-1 dark matter, there are two such dimension-6 operators:
\begin{eqnarray}
{\cal O} &=& {1 \over 2 \Lambda^2} B_{\{ \mu} B_{\nu \}} \left(\bar f \gamma^{\{ \mu} \partial^{\nu \} }  f
- \partial^{\{ \nu} \bar f \gamma^{\mu \} }  f\right)\,,
\nonumber\\
{\cal O'} &=& {1 \over 2 \Lambda^2} B_{\{ \mu} B_{\nu \}} \left(\bar f \gamma^{\{ \mu} \gamma^5 \partial^{\nu \} } f
- \partial^{\{ \nu} \bar f \gamma^{\mu \} } \gamma^5 f\right)\,.
\end{eqnarray}
These operators (which were not discussed in Refs.~\cite{Kumar2013,Kumar:2015wya}) yield a nontrivial matrix element between an $s$-wave $J=2$ dark matter initial state and an
$\bar f f$ final state with no Weyl spinor mixing.  As these operators are both $CP$-even and respect SM
flavor symmetries, they cannot be projected out.

We can verify this result with explicit calculation.  
Henceforth, we assume $m_f=0$.
We consider a model in which the spin-1 dark particle $B$ couples
to a SM fermion $f$ through exchange of a heavy charged fermion $\Psi$ via the interaction Lagrangian,
\begin{equation}
  \mathcal{L} = \lambda_L \bar \Psi \gamma^\mu P_L f B_\mu + \lambda^*_L \bar f \gamma^\mu P_L \Psi B_\mu\,,
\end{equation}
where $\lambda_{L}$ is a dimensionless coupling and we have assumed that the dark sector only couples to
left-handed SM fermions.
% The amplitude for the $t$-channel annihilation process $B(p_1)B(p_2)\rightarrow f(k_1)\bar{f}(k_2)$  is given by
% the Feynman diagrams in Fig.~\ref{Fig:2badm}:
% \begin{figure}[!htbp]
% \begin{center}
%   \subfloat[\label{sf:ggh1}]{
%    \includegraphics[scale=0.4]{vdmat}}~~
%   \subfloat[\label{sf:yyh1}]{
%    \includegraphics[scale=0.4]{vdmau}}
%    \caption{Feynman diagram for $B(p_1)B(p_2)\rightarrow f(k_1)\bar{f}(k_2)$ process. Directions of momentum and fermion
% number flow are shown in the diagrams.}
% \label{Fig:2badm}
% \end{center}
% \end{figure}
The unpolarized annihilation cross section for the $t$-channel annihilation process $B(p_1)B(p_2)\rightarrow f(k_1)\bar{f}(k_2)$
% for this model 
has been computed in Refs.~\cite{Cheng2002, Farzan2012, Yu2014}, and indeed it is nonvanishing in the nonrelativistic limit.

The $L=0$, $S=0$, $J=0$ initial state can be written in the individual spin basis as
\begin{eqnarray}
|J=0, J_z =0 \rangle &=& \frac{1}{\sqrt{3}}|S_z^1 =+1, S_z^2=-1 \rangle
-\frac{1}{\sqrt{3}}|S_z^1 =0, S_z^2=0 \rangle
+\frac{1}{\sqrt{3}}|S_z^1 =-1, S_z^2=+1 \rangle\,,
\end{eqnarray}
and one can verify that the matrix element for the annihilation of this state to $\bar f f$ 
vanishes in the nonrelativistic limit.  Similarly the matrix element for annihilation of the
$J=1$ initial state also vanishes in the nonrelativistic limit; the two diagrams cancel, indicative of
the fact that two identical particles cannot be in an $L=0$, $S=1$, $J=1$ initial state.  But the
matrix element for annihilation of the $J=2$ initial state is nonvanishing.

\section{Internal Bremsstrahlung}

We now consider the matrix element for the annihilation of the $J=0$ initial state to the $\bar f f \gamma$
final state.  
% The Feynman diagrams contributing to the internal bremsstrahlung process are shown in fig.(\ref{fig:ib}).

% \begin{figure}[!htbp]
% \begin{center}
%   \subfloat[\label{fig:t1}]{
%    \includegraphics[scale=0.4]{VDM_VIB_Fig1t_new}}~~
%   \subfloat[\label{fig:u1}]{
%    \includegraphics[scale=0.4]{VDM_VIB_Fig1u_new}}\\
%      \subfloat[\label{fig:t2}]{
%    \includegraphics[scale=0.4]{VDM_VIB_Fig2t_new}}~~
%   \subfloat[\label{fig:u2}]{
%    \includegraphics[scale=0.4]{VDM_VIB_Fig2u_new}}\\
%    \subfloat[\label{fig:u3}]{
%    \includegraphics[scale=0.4]{VDM_VIB_Fig3t_new}}~~
%   \subfloat[\label{fig:t3}]{
%    \includegraphics[scale=0.4]{VDM_VIB_Fig3u_new}}
%    \caption{Feynman diagrams contributing to internal bremsstrahlung for vector dark matter. Directions of
% momentum and fermion number flow are shown in the diagrams.}
% \label{fig:ib}
% \end{center}
% \end{figure}
%

The squared amplitude for annihilation of the $J=0$ initial state (summed over final state spins) is
\begin{equation}
\label{eq:amp3bodyS}
\sum_{spins} |\M_{J=0}|^2=\frac{32\pi\alpha\lambda^4_L(2+r^2)^2 }{ 3 m_B^2}\, \frac{ 4  (1 - x)(2 + 2 y^2 + 2 y( x-2) - 2 x+ x^2)}{
 (1 - r^2 - 2 y)^2 (3 + r^2- 2 x - 2 y)^2}\,,
\end{equation}
where $r \equiv m_{\Psi}/m_B$.  Here $x \equiv 2E_\gamma/\sqrt{s}$,
$y \equiv 2E_f/\sqrt{s}$, and $\sqrt{s}$ is the center-of-mass energy.  The reduced energy parameters $x$ and $y$ are subject to the kinematic
constraints $0 \leq x,y, \leq 1$, $x+y \geq 1$.

\subsection{Relation to the spin-0 and spin-1/2 cases}

Note that the $x$ and $y$ dependence of Eq.~(\ref{eq:amp3bodyS}) is identical to that for the cases of spin-0 and spin-1/2 dark matter~\cite{Barger:2009xe,Barger2012}.  An explanation for the identicalness
of the internal bremsstrhalung spectra for spin-0 and spin-1/2 dark matter that relies on effective operators was put forward in
Ref.~\cite{Barger2012}, and extended in Ref.~\cite{Giacchino:2014moa} using the operator classification of Ref.~\cite{DeSimone:2013gj}.
 It has been shown that in the heavy mediator limit ($r \gg 1$), the effective 5-point contact operators that lead to internal bremsstrahlung must be of dimension $\geq 8$~\cite{Barger2012, DeSimone:2013gj}.  The dominant
operators are thus dimension 8: there are 5 such operators for Majorana fermion dark matter, and 7 such operators for
real scalar dark matter~\cite{DeSimone:2013gj}.  But it turns out that all of these operators produce identical photon spectra; although any particular
model will be realized as one particular linear combination of these operators, the resulting photon spectrum is
necessarily universal.    If one does not take the limit $r \gg 1$, the only change to the form of the amplitude arises from the denominators of
the propagators of the heavy mediators~\cite{Giacchino:2014moa}. But as the annihilation process is $t$-channel for both spin-0 and spin-1/2 dark matter, the
denominators of the propagators in these two cases are the same, implying that the spectrum remains universal even outside
of the contact operator limit.\footnote{This generalization fails if internal bremsstrahlung can occur from an $s$-channel
diagram, as is the case with internal Higgsstrahlung~\cite{Kumar:2016mrq}.}

This argument easily generalizes to the case of spin-1 dark matter.  Each of the contact operators
can be written as a dark matter bilinear with some Lorentz tensor structure, contracted with a SM trilinear
with the same Lorentz structure, provided that the SM factor can produce an $\bar f f \gamma$ final state and that the
DM factor has a nontrivial matrix element with an $s$-wave initial state of identical particles.
The shape of the photon spectrum is determined only by the SM trilinear factor because
the DM bilinear contributes a constant factor to the matrix element which is independent of $x$ and $y$.

For the spin-0 case, the DM bilinear is necessarily either a 0-index or 2-index tensor, while in the spin-1/2
case the DM bilinear is a 1-index tensor~\cite{DeSimone:2013gj}.  For the case of spin-1 dark matter, the only bilinears
that have a nontrivial matrix element with the $J=0$ $s$-wave initial state are $B_\mu B^\mu$ and $B_\mu B_\nu$.  Since
the initial state has $J=0$, only a rotationally invariant piece of the tensor can contribute; for the bilinear $B_\mu B_\nu$, this
piece scales as $\delta_{ii}$ in the nonrelativistic limit.  The contact operators for internal bremsstrahlung of spin-1 dark matter can
therefore be written in terms of the operators for spin-0 dark matter, found in Ref.~\cite{DeSimone:2013gj}, by the replacements
$\phi^2 \rightarrow B_\mu B^\mu$, $\partial_\mu \phi \partial_\nu \phi \rightarrow B_\mu B_\nu$.  Because the photon spectrum
is determined by the SM factor, these replacements do not alter the spectrum, which is thus identical for the case of spin-1
and spin-0 dark matter. 

There is one subtlety to this argument.  For the case of spin-0 dark matter, the bilinear $\partial_\mu \phi
\partial_\nu \phi$ has a nontrivial matrix element with the $J=0$ initial state only if $\mu = \nu =0$, so only the corresponding
terms in the SM factor contribute to the photon spectrum.  But with the replacement $\partial_\mu \phi \partial_\nu \phi \rightarrow B_\mu B_\nu$,
only the terms in the DM bilinear with $\mu = \nu =i$ are nontrivial, implying that a different set of terms in the SM trilinear contribute
to the photon spectrum for the spin-1 case.  
The SM trilinear with which the 2-index DM bilinear is contracted is
$\bar f \gamma^\mu \overrightarrow{D}^\nu P_L f - \bar f \overleftarrow{D}^\nu \gamma^\mu  P_L f$~\cite{DeSimone:2013gj}, where $D$ is the SM covariant derivative. Since the matrix element for this
factor vanishes if contracted with $g_{\mu \nu}$~\cite{DeSimone:2013gj}, the contributions from the SM trilinear corresponding to $\mu = \nu =i$ and $\mu = \nu =0$ are identical, thereby implying that the contribution to the matrix element for the spin-0 case is the same
as for the spin-1 case.

There is one last interesting feature to note from this construction.  The replacement $\partial_\mu \phi \partial_\nu \phi \rightarrow B_\mu B_\nu$,
gives the effective operator $B_\mu B_\nu (\bar f \gamma^\mu \overrightarrow{D}^\nu P_L f - \bar f \overleftarrow{D}^\nu \gamma^\mu  P_L f)$, 
which is dimension-6.  So, this operator can provide a contribution to the squared matrix element which scales as $r^{-4}$, instead of the $r^{-8}$ scaling which
necessarily appears for spin-0 or spin-1/2 dark matter.  Indeed, we see this scaling in Eq.~(\ref{eq:amp3bodyS}).
This implies that in the heavy mediator limit, there is less suppression of the internal
bremsstrahlung cross section for the case of spin-1 dark matter.

\subsection{Realization of internal bremsstrahlung as the dominant annihilation channel}
%%%%%%%%%%%%%%%%%%%%%%%%%%%%%%%%%%%%%%%%%%%%%%%%%%%%%%%%%%%%%%%%%%%%%%%%%%%%%%%%%%%%%%%%%%%%%%%%%%%%%%%%%%%%%%%%

As we have seen, the internal bremsstrahlung annihilation process can only be significant if annihilation from the
$J=2$ initial dark matter state is suppressed.  So internal bremsstrahlung is only relevant if the
initial state is polarized.  A simple scenario in which this can happen is if the dark matter particles predominantly annihilate by
first forming a nonrelativistic $BB$ bound state, which then decays to its ground state, before the constituents finally annihilate.
If the ground state is $J=2$, then $s$-wave annihilation to a two-body final state will dominate, while if the
ground state is $J=0$, then $s$-wave annihilation to a three-body $\bar f f \gamma$ final state will dominate.  If
the ground state is $J=1$, then $p$-wave annihilation to a two-body final state will dominate.

There is a large body of work on self-interacting dark matter, including models in which dark sector particles
form composite bound states; see, for example, Refs.~\cite{Feng:2009mn, Alves:2009nf, Kaplan:2009de}. A detailed formulation of the confining potential, and the resulting spectroscopy, is beyond
the scope of this work.   We note, however, that the dynamics which generate the confining potential are independent of the
field $\Psi$ that mediates the interaction between the dark matter and the SM.  Thus, $m_\Psi$ can be much larger
than the scale of the confining potential, and the lifetime of the bound state can easily be large compared to the timescale on which it 
deexcites to the ground state.

There are two distinct classes of models within this scenario: dark matter may largely
consist of such bound states, or the dark matter particles may largely be unbound, with the formation of bound states
followed relatively quickly by annihilation of the constituents.  In the former case, one may just as well treat the
$J=0$ bound state as a composite spin-0 dark matter particle, whose two-body decays to SM fermions are chirality suppressed~\cite{Barger:2009xe}.  In
the latter case, the spin-1 particles $B$ in fact constitute the dark matter.  This distinction can be significant
for the purposes of direct detection.  But in either case, the doubly-differential decay rate of the bound state  may be
written as
\begin{eqnarray}
{d^2 \Gamma \over dx dy} &=&\frac{|\phi(0)|^2}{128 \pi^3} \left( \sum_{spins} |\M_{J=0}|^2 \right),
\end{eqnarray}
where $\phi(0)$ is the wavefunction of the bound state evaluated at the origin.
%%%%%%%%%%%%%%%%%%%%%%%%%%%%%%%%%%%%%%%%%%%%%%%%%%%%%%%%%%%%%%%%%%%%%%%%%%%%%%%%%%%%%%%%%%%%%%%%%%%%%%%%%%%%%%%%%%%
\section{Conclusions}
%%%%%%%%%%%%%%%%%%%%%%%%%%%%%%%%%%%%%%%%%%%%%%%%%%%%%%%%%%%%%%%%%%%%%%%%%%%%%%%%%%%%%%%%%%%%%%%%%%%%%%%%%%%%%%%%%%%%

We considered the annihilation of self-conjugate spin-1 dark matter to Standard Model fermions, assuming that
 flavor violation is minimal.  We have shown that the $BB \rightarrow \bar f f$
annihilation cross section can be chirality suppressed, but only if the initial state is $J=0$; the $J=2$ state
has an unsuppressed $2 \rightarrow 2$ annihilation cross section.  For the $J=0$ state, the dominant $s$-wave
annihilation process yields a three-body final state $\bar f f \gamma$ via internal bremsstrahlung,
with a spectrum that is identical to the case of self-conjugate spin-0 or spin-1/2 dark matter.

The typical suppression of the internal bremsstrahlung cross section by a factor of the fine structure constant implies that this process is unimportant
for the case in which the initial DM state is unpolarized.  But there are scenarios for which the DM state
is polarized, as in the context of self-interacting dark matter.  In particular, if dark matter forms a
nonrelativistic $BB$ bound state, which decays to a $J=0$ ground state before the constituents annihilate, then internal bremsstrahlung
could be the dominant annihilation process.

We have shown that for spin-1 dark matter, the shape of the photon spectrum arising from internal bremsstrahlung is necessarily
the same as for spin-0 and spin-1/2 dark matter, generalizing previous arguments regarding the universality of the photon spectrum.  But
unlike the spin-0 or spin-1/2 cases, in which the annihilation matrix element is suppressed by $m_\Psi^{-4}$ in the heavy
mediator limit, in the spin-1 case the matrix element is only suppressed by $m_\Psi^{-2}$.  This is particularly interesting
because of its impact on complementary searches at the LHC.  In order for internal bremsstrahlung to be possible, the mediator
$\Psi$ must be charged, and there are tight constraints on new charged particles from collider experiments.  If $m_\Psi$ is increased in order
to evade those constraints, then the internal bremsstrahlung cross section becomes heavily suppressed.  As a result, many studies of
the IB photon spectrum have necessarily focused on the regime where the dark matter and the mediator are nearly degenerate; in this
region of parameter space, the internal bremsstrahlung cross section is not heavily suppressed, and the mediator escapes collider
searches because the jets/leptons produced by its decay are soft. For the case of spin-1 dark matter, however, $r$ can be made much
larger without heavily suppressing the annihilation cross section.  This opens a new window in parameter space in which one can search
for dark matter annihilation via internal bremsstrahlung.

\section{Acknowledgements}
The  authors  acknowledge  the hospitality provided  by the  WHEPP-XIV workshop held at IIT Kanpur, India, during which this work was initiated. G.B. and G.T. thank S. Mohanty for helpful discussions.
J.K. is supported in part by NSF CAREER grant PHY-1250573.  D.M. is supported in part by
DOE grant DE-SC0010504.
%%%%%%%%%%%%%%%%%%%%%%%%%%%%%%%%%%%%%%%%%%%%%%%%%%%%%%%%%%%%%%%%%%%%%%%%%%%%%%%%%%%%%%%%%%%%%%%%%%%%%%%%%%%%%%%%%%%%%

%%%%%%%%%%%     Bibliography     %%%%%%%%%%%%%%%%
%%%%%%%%%%%%%%%%%%%%%%%%%%%%%%%%%%%%%%%%%%%%%%%%%
%%%%%%%%%%%%%%%%%%%%%%%%%%%%%%%%%%%%%%%%%%%%%%%%%%%%%%%%%%%%%%%%%%%%%%%%%%%%%%%%%%%%%%%%%%%%%%%%%%%%%%%%%%%%%%%%%%%
% \bibliographystyle{plain}
\bibliography{vdmref}
%%%%%%%%%%%%%%%%%%%%%%%%%%%%%%%%%%%%%%%%%%%%%%%%%%%%%%%%%%%%%%%%%%%%%%%%%%%%%%%%%%%%%%%%%%%%%%%%%%%%%%%%%%%%%%%%%%%
\end{document}